\documentclass[12pt]{iopart} 
\usepackage{iopams}  
\usepackage{setstack}    
\usepackage[pdftex]{graphicx,color}    
\usepackage{tensor}
\usepackage{hyperref}

\newcommand{\LW}{Li\'enard--Wiechert}  
\newcommand{\be}{\begin{equation}}
\newcommand{\ee}{\end{equation}}
\newcommand{\bea}{\begin{eqnarray}}
\newcommand{\eea}{\end{eqnarray}}
\newcommand{\diag}{{\rm diag}}

\newcommand{\N}[1]{\tensor{N}{#1}}
\newcommand{\kap}[1]{\tensor{\kappa}{#1}}

\newcommand{\mB}{m_{{}_{\!B\!}}}     
\newcommand{\ret}{{\rm ret}}
\newcommand{\adv}{{\rm adv}}

\begin{document}

\title[Self-force and mass renormalization]{The alternative to classical mass renormalization for tube-based self-force calculations}

\author{Andrew H Norton}


\address{Department of Mathematical Sciences, 
and CUDOS\footnote{Centre for Ultrahigh bandwidth Devices for Optical Systems.}, 
University of Technology Sydney, Australia.}

\ead{andrew.norton@aei.mpg.de}

\medskip
\address{Published 29 April 2009: Class. Quantum Grav. \href{http://dx.doi.org/10.1088/0264-9381/26/10/105009}{{\bf 26} (2009) 105009}}


\begin{abstract}
To date, classical mass renormalization has been invoked in all tube-based self-force calculations,
thus following the method introduced in Dirac's 1938 calculation of the electromagnetic self-force
for the classical radiating electron.
In this paper a new tube method is described that does not rely on a mass renormalization procedure.
As a result, exact self-force calculations become possible for classical radiating systems of finite size.
A new derivation of the Lorentz--Dirac equation is given and the relationship between the new tube method and the
classical mass renormalization procedure is explained. It is
expected that a similar tube method could be used to obtain rigorous results in the
gravitational self-force problem.
\end{abstract}
%
%
%
%
%

%

\section{Introduction}

The classical mass renormalization procedure was introduced 70 years ago by Dirac~\cite{Dirac:38} and
has since featured in all tube-based calculations of self-forces on radiating particles. 
This includes not only calculations involving the electromagnetic self-force on a radiating charge in Minkowski spacetime, 
as exemplified by 
\cite{Dirac:38}--\cite{Poisson:99}, 
but also electromagnetic self-force calculations for a charged particle in a curved background spacetime 
\cite{dewitt60,Hobbs:68}, and the tube-based approach to the gravitational self-force problem, as taken 
in \cite{Mino:97} and reviewed in 
\cite{Poisson:04}--\cite{Mino:05}. 
Nevertheless, the mathematical status
of the classical mass renormalization procedure has remained unclear and justification for its use has principally been that 
it has given results that agree with other self-force methods. As recently emphasised in \cite{Gralla:08}, this is 
particularly true of the gravitational self-force problem for which the concept of a point particle makes even less sense 
than in the electromagnetic case. 

This paper describes a tube method for determining the self-force on a radiating particle or system of finite non-zero size.
Dirac's tube method is extended so that the calculations go through with a worldtube that has, at all stages of the calculation, 
a non-zero radius. The physical picture that emerges is very different to that which accompanies the classical mass 
renormalization procedure. In retrospect, one finds that the renormalization procedure is most easily understood
as a short-cut that involves assuming, rather than deriving, a certain intermediate result 
(our equation (\ref{PcapSeries})) that is required for a proper derivation of the self-force.

The new tube method is presented here for the problem of electromagnetic self-force in Minkowski spacetime and,  
as an example, a derivation of the Lorentz--Dirac equation is given that makes no mention of mass renormalization. 
Since the method is based on quite general
considerations of energy-momentum conservation, the same approach ought to be applicable in any radiation 
dynamics problem for which a well defined self-force exists. In particular, it seems likely that a similar
tube method could be used to obtain rigorous results in the gravitational self-force problem.  

\subsection{Background}

Dirac's tube method for calculating the self-force on a radiating particle was used in \cite{Dirac:38} 
to derive what is now known as the Lorentz--Dirac equation of motion for a classical 
radiating particle of charge $q$ and mass $m$ in a background electromagnetic field $F$,
\be
   mc \,\ddot z^\alpha \;=\; q \tensor{F}{^\alpha_\beta} \dot z^\beta 
                 +  \frac{q^2}{4\pi \epsilon_0 c}\,\frac{2}{3}\left( \delta^\alpha_\beta +  \dot z^\alpha  \dot z_\beta\right) \dddot z^\beta\,,
\label{dirac_eom1}
\ee 
where the worldline has equation $x = z(s)$ in Minkowski coordinates, overdots denote differentiation with respect 
to $s$, and the parameterization is such that $\dot z^\alpha \dot z_\alpha=-1$.

The calculation that led Dirac to (\ref{dirac_eom1}) can be described as
follows. One starts by deriving a balance
relation that follows from energy-momentum conservation, as applied to
the electromagnetic field within a worldtube containing the radiating
particle.  The balance relation expresses the fact that the rate of
change of 4-momentum flowing within the worldtube must equal the rate
at which electromagnetic energy-momentum is radiated through the wall
of the tube. The derivation of Dirac's balance relation is lengthy but
straightforward. It is easily found (for a given worldtube) using computer algebra and takes 
the form of a power series in the tube radius $r$, which is taken to be small compared with the 
radius of curvature of the particle worldline (the inverse of the particle acceleration).
  
The worldtube exists only to facilitate the calculations, so the tube radius can not enter into the 
sought-after equation of motion. Dirac's approach for eliminating the tube radius was to let
$r\rightarrow 0$, in which case only quantities determined by the particle worldline
can appear in the final result. The difficulty with this idea is 
that because the particle must remain enclosed within the worldtube, it must then
be modelled as being point-like. The electromagnetic field is then singular 
on the particle worldline, causing the required $r\rightarrow 0$ limit to be neither defined
nor physically sensible.  In fact, to obtain the desired result
(\ref{dirac_eom1}) using this approach, one must formally subtract out the divergent
Coulomb field energy of the point charge by assigning the particle an
infinite negative bare mass --- the procedure that has come to be known as
classical mass renormalization. 

To avoid the absurdities of classical mass renormalization, one needs
a way to eliminate the tube radius from Dirac's balance relation
without taking $r\rightarrow 0$. Moreover, simply taking $r$ to be
small and neglecting $\Or(r)$ terms does not quite work, as this would result in a finite mass renormalization
that is $r$-dependent. In effect, one would be saying that the field-energy contribution to the 
mass of the particle somehow depended on the radius of an imaginary tube that was introduced purely as
a mathematical convenience.

In this paper we shall see how to cancel all of the $r$-dependent terms from the Dirac balance
relation for any classical radiating system. This is done by making
better use of the information available in the problem, including the
fact that the background electromagnetic field must satisfy the
source-free Maxwell equations throughout the spacetime region occupied
by the radiating particle or system. The worldtube used in the
calculations will have a non-zero radius and there will be no need to
assume any mass renormalization, finite or otherwise.

\subsection{Organisation of the paper}

The main ideas and results are presented in Section~\ref{tube_method}, where the tube method is described with reference
to the Lorentz--Dirac example, that is, for the problem of determining the self-force for a charged particle in Minkowski spacetime. 
As far as possible, all distracting calculations have been deferred to later 
sections. In particular, the geometry of the worldtube is dealt with in Section~\ref{tube_coordinates} and the electromagnetic
field calculations appear in Section~\ref{stress_energy_tensor}. A {\em Mathematica} notebook for all of the 
calculations is available as an electronic supplement \cite{Norton:08}.

In principle, the worldtube could contain
any classical radiating system, so it should be kept
in mind that the only parts of Section~\ref{tube_method} that are truly specific to 
the Lorentz--Dirac example are the values quoted for the coefficients in the two series expansions 
of $\partial_s P_{\rm cap}(s,r)$ (equations (\ref{DiracBalance}) and (\ref{secondexp})). 

The unit system used in the paper is the SI and the Minkowski metric is taken to be $\eta = \diag(-1,1,1,1)$. Thus, any equation 
here can be converted to the geometrized Gaussian CGS units used by Hawking and Ellis \cite{HawkingEllis:73}, 
Misner, Thorne and Wheeler \cite{MTW:73} and Wald \cite{Wald:84}, for example, by setting $c = G = 1$ 
and $4\pi\epsilon_0 = 1$ (hence also $\mu_0 = 4\pi$).

\section{The tube method}
\label{tube_method} 

We start by supposing the particle has a physically sensible (but unknown) classical structure, for example, 
as a soliton in some nonlinear field theory that reduces to Maxwell electrodynamics in its weak-field 
limit, or perhaps as some classical electron model \cite{pearle82}. Here, physically sensible is taken to mean there
exists a conserved stress-energy tensor $T$ for this underlying theory, and that the particle has a finite size.
Thus, we suppose
\bea
                    \nabla \cdot T = 0 \,, &
\label{claw}
\\
                     T = \left\{ 
\begin{array}{ll} 
                              T_{\rm matter}  \qquad  & {\rm for\ } r \le r_0\,,
\\
                              T_{\rm e.m.}            & {\rm otherwise}\,,
\end{array} \right.
\label{Tdefn}
\eea 
where $T_{\rm e.m.}$ is a purely electromagnetic (vacuum) stress-energy tensor and the radius $r_0$ defines, for some predetermined 
level of accuracy, the extent of the matter fields or size of the particle structure.
A precise value for this structure radius is not needed. It will suffice to assume that 
such an $r_0$ exists, so that the particle can be considered to lie within a worldtube of radius $r_0$.
The worldline ${\cal C}$ on which this tube is centred will serve as a reference worldline for the particle,
about which the asymptotic field of the particle shall be prescribed, usually in terms of a multipole expansion 
\cite{vanWeert:74,Teitelboim:80,ellis66,vanWeert:73}.

The simplest such prescription is that the charged
particle have an asymptotic field that exactly coincides with that of an electric monopole, as
given by the retarded {\LW} potential for a point-source with worldline ${\cal C}$. Later, we shall see 
that this case leads directly to the Lorentz--Dirac equation (\ref{dirac_eom1}). More generally, the asymptotic field of 
the (non-zero size) charged particle would include higher order multipoles, and the corresponding self-force would differ 
from the Lorentz--Dirac result.

The worldtube for which Dirac's  
balance relation is to be derived will be chosen to have a tube radius $r$ that is much larger than the structure radius $r_0$, 
so that it lies in the asymptotic field of the particle, where the prescribed multipole expansion
can be used. On the other hand, we still take $r$ to be small compared with the scale set by the radius of curvature 
of ${\cal C}$, that is $|r \kappa_1| \ll 1$, where $\kappa_1$ (defined by (\ref{dot_N_0})) is the first Frenet--Serret 
curvature of ${\cal C}$. The situation described is shown in Figure~\ref{tubefig}. 

\begin{figure}[h!tb]
        \centering
        \resizebox{14cm}{!}{\input norton_figure1.tex}
        \caption{The worldtube $\Sigma_{\rm tube}(s,r)$ used in the self-force calculations. The tube is of length $s$ and 
        has radius $r \gg r_0$
        where $r_0$ is the structure radius of the particle. The momentum $P_{\rm cap}(s,r)$ includes unknown contributions 
        from the stress-energy tensor associated with the particle structure. Nevertheless, expressions for the derivatives 
        $\partial_s P_{\rm tube}(s,r)$ and $\partial_r P_{\rm cap}(s,r)$ can still be evaluated
        in the asymptotic region $r \gg r_0$, and the self-force on the radiating particle thereby calculated using
        conservation of energy-momentum over the 4-volume $V(s,r)$.} 
        \label{tubefig}
\end{figure}

Let ${\cal C}$ be a smooth timelike worldline with equation $x = z(s)$ where $x^\alpha$, $\alpha=0,\ldots,3$, 
are Minkowski coordinates with metric $\eta = \diag(-1,1,1,1)$, and the parameter $s$ is proper 
distance ($c \; \times$ proper time) 
along ${\cal C}$, so that $\dot z \cdot \dot z = -1$. 
In Section~\ref{tube_coordinates}, spherical polar tube coordinates $\{s,\,r,\,\vartheta,\,\varphi\}$ are 
constructed using the Frenet frame of
${\cal C}$, such that the worldline parameter $s$ coincides on ${\cal C}$ with the spacetime coordinate $s$, and ${\cal C}$ 
has the equation $r=0$. The various integration surfaces that are needed can then be specified in terms of coordinate surfaces.   
The worldtube $\Sigma_{\rm tube}(s,r)$, the tube-cap $\Sigma_{\rm cap}(s,r)$,  
their intersection 2-sphere $\Omega(s,r)$, and the 4-volume $V(s,r)$ are defined by
\bea
            \Sigma_{\rm tube}(\bar s, \bar r) \;=\; \left\{s \in [0,\,\bar s],\, r = \bar r\right\},
\label{tube_defn}
\\
            \Sigma_{\rm cap}(\bar s, \bar r) \;=\; \left\{s = \bar s,\, r \in [0,\, \bar r]\right\},     
\label{cap_defn}
\\
            \Omega(\bar s, \bar r) \;=\; \left\{s = \bar s,\, r = \bar r\right\},     
\label{s2_defn}
\\
            V(\bar s, \bar r) \;=\; \left\{s \in [0,\,\bar s],\, r \in [0,\, \bar r]\right\},    
\label{volume_defn}
\eea
where in each case the angular coordinates assume the range $(\vartheta,\varphi) \in [0,\,\pi]\times [0,\,2\pi)$.
Figure~\ref{tubefig} shows these various structures for the time-symmetric tube coordinate system 
defined in Section~\ref{symmetric_tube_coords}. For this coordinate system the worldtube is of Dirac-type, meaning that
the tube caps are orthogonal 
to ${\cal C}$, as used in \cite{Dirac:38}. The calculations can also, for example, be based on retarded (advanced) tube coordinates, in which 
case the tube caps are parts of forward (backward) null cones, as described in Section~\ref{ret_tube_coords}. The results quoted in the 
present section are for the Dirac-type worldtube of Figure~\ref{tubefig}.

Let $P_{\rm cap}(s,r)$ be the momentum flux through $\Sigma_{\rm cap}(s,r)$, and 
let $P_{\rm tube}(s,r)$ be the (outward) momentum flux across the worldtube $\Sigma_{\rm tube}(s,r)$. 
In Section~\ref{conservation} it is shown that
\bea
           P_{\rm cap}(s,r) \;=\; \frac{1}{c}\int_{r^\prime = 0}^r  \int_{\Omega(s,r^\prime)}
                                                   T\cdot \nu_{\rm cap}\,\rmd\Omega\,\rmd r^\prime\,,
\label{Pcap}
\\
           P_{\rm tube}(s,r) \;=\; \frac{1}{c}\int_{s^\prime = 0}^s  \int_{\Omega(s^\prime,r)} 
                                                   T\cdot \nu_{\rm tube}\,\rmd \Omega\,\rmd s^\prime\,,               
\label{Ptube}
\eea
where $\rmd\Omega$ is the element of area for the 2-sphere $\Omega(r,s)$, 
and the vectors $\nu_{\rm cap}$ and  $\nu_{\rm tube}$ (given by (\ref{nu_cap}) and (\ref{nu_tube})) are normals for the 
integration 3-surfaces. 
Observe that expressions (\ref{Pcap}) and (\ref{Ptube}) depend on $r$ and $s$ respectively, 
only through the upper limits of the outer integrals. Differentiating (\ref{Pcap}) with respect to $r$ and (\ref{Ptube}) 
with respect to $s$ therefore gives
\bea
      \frac{\partial P_{\rm cap}(s,r)}{\partial r} \;=\;  \frac{1}{c} \int_{\Omega(s,r)} T\cdot \nu_{\rm cap}\,\rmd\Omega\,,
\label{PcapDr}
\\
      \frac{\partial P_{\rm tube}(s,r)}{\partial s} \;=\; \frac{1}{c}  \int_{\Omega(s,r)} T\cdot \nu_{\rm tube}\,\rmd\Omega\,.
\label{PtubeDs}
\eea
Following Dirac, equation (\ref{PtubeDs}) becomes a relation involving  $P_{\rm cap}(s,r)$ by considering
energy-momentum conservation for the fields within the 4-volume $V(s, r)$. Dirac assumes 
these fields are purely electromagnetic, so that $T = T_{\rm e.m.}$ all the way up to the worldline ${\cal C}$, but in our
case only the asymptotic field of the particle is known to us. The particle structure is unknown, as is the 
stress-energy tensor $T_{\rm matter}$ associated with this structure. Nevertheless, Gauss' theorem still applies and 
in Section \ref{gauss_formula} it is shown that by integrating (\ref{claw}) over the 4-volume $V(s, r)$ one has
\be
              P_{\rm cap}(s,r) \;=\; P_{\rm cap}(0,r) - P_{\rm tube}(s,r)\,.
\label{plaw}
\ee
Differentiating (\ref{plaw}) with respect to $s$ and making use of (\ref{PtubeDs}) then gives 
\be
       \frac{\partial P_{\rm cap}(s,r)}{\partial s} \;=\;  - \,\frac{1}{c} \int_{\Omega(s,r)} T\cdot \nu_{\rm tube}\,\rmd\Omega\,.
\label{int1}
\ee
We now use the fact that the worldtube has radius $r \gg r_0$. The 2-sphere $\Omega(s,r)$ over 
which the integral (\ref{int1}) is evaluated therefore lies in the asymptotic field of the particle, where the stress-energy tensor
is of known electromagnetic form. After replacing $T$ in (\ref{int1}) by $T_{\rm e.m.}$, one then expands the integrand 
$T_{\rm e.m.} \!\cdot \nu_{\rm tube}$, as a power series in the tube radius $r$, which although large with respect to the 
structure radius $r_0$, is still assumed to be small with respect to the 
scale set by the worldline curvature ($|r \kappa_1| \ll 1$). The angular integrations then involve trigonometric polynomials 
in $\{\vartheta,\varphi\}$ and may be evaluated exactly. 
In Section~\ref{stress_energy_tensor} these calculations are described in further detail
for the Lorentz--Dirac example, in which case $T_{\rm e.m.}$ is taken to be the stress-energy tensor of an electromagnetic 
field that is the 
sum of a  background field $F$ and an electric monopole (retarded {\LW}) field for the particle.
The result is Dirac's balance relation (equation (18) in \cite{Dirac:38}),
\be
       \frac{\partial P_{\rm cap}(s,r)}{\partial s}
                    \;=\;  - \,\frac{1}{c}\int_{\Omega(s,r)} T_{\rm e.m.}\!\cdot \nu_{\rm tube}\,\rmd\Omega
                    \;=\;  \sum_{k = -1}^\infty a_k(s)\, r^k\,,
\label{DiracBalance}
\ee
where the first few coefficients in the series are found to be \cite{Norton:08},
\bea
          a_{-1}(s) &=&\, \frac{-q^2}{8 \pi\epsilon_0 c}\, \kappa_1 N_1\,,
\\
          a_0(s) &=&\, q F\cdot N_0 +  \frac{q^2}{6 \pi\epsilon_0 c} 
                                     \left(\tensor{\dot \kappa}{_1} \N{_1} + \kap{_1}\kap{_2}\N{_2} \right) \,,
\label{a0coeff}
\\
          a_1(s) &=&\, \frac{q^2}{48 \pi\epsilon_0 c} \Bigl( 2 \kap{_1} \tensor{\dot \kappa}{_1} \N{_0}
                           + \left( 3 \kap{_1}^3 + 4 \kap{_1} \kap{_2}^2 - 4 \tensor{\ddot \kappa}{_1}\right) \N{_1}
\nonumber \\ && {}
                           - 4 \left( \kap{_1} \tensor{\dot \kappa}{_2} + 2 \kap{_2} \tensor{\dot \kappa}{_1} \right) \N{_2}
                           - 4 \kap{_1}\kap{_2}\kap{_3} \N{_3} \Bigr)\,.          
\label{a1coeff}
\eea
Here $\N{_i}$ and $\kap{_a}$ are respectively the Frenet--Serret frame vectors and curvatures of ${\cal C}$ 
(see Section~\ref{frenet_frame} for definitions).
For $k \ge 2$, the coefficients $a_k(s)$ involve successively higher derivatives of $F$ evaluated on ${\cal C}$, and become 
rapidly more complicated. 

A similar calculation is used to evaluate (\ref{PcapDr}). Again, on noting that $\Omega(s,r)$ lies in the 
asymptotic field of the particle we replace $T$ by $T_{\rm e.m.}$, expand the integrand as a power series in $r$
and then do the angular integrations. The result is
\be
       \frac{\partial P_{\rm cap}(s,r)}{\partial r}
                    \;=\;  \frac{1}{c}\int_{\Omega(s,r)} T_{\rm e.m.}\!\cdot \nu_{\rm cap}\,\rmd\Omega
                    \;=\;  \sum_{k = -2}^\infty b_k(s)\, r^k\,,
\label{PcapDrSeries}
\ee
where the first few coefficients in the series are
\bea
             b_{-2}(s)  &=&  \frac{q^2}{8 \pi\epsilon_0 c}\, N_0\,,
\label{b_2}
\\
             b_{-1}(s)  &=&  0\,,
\\
             b_{0}(s)  &=&  \frac{q^2}{48 \pi\epsilon_0 c} \left( 3 \kap{_1}^2 \N{_0} 
                                                                - 4 \tensor{\dot \kappa}{_1} \N{_1} 
                                                                -4 \kap{_1}\kap{_2} \N{_2} \right) \,.    
\label{b_0}
\eea
Assuming convergence of the series, we now integrate (\ref{PcapDrSeries}) as a partial differential equation in 
independent variables $\{s,r\}$ to get,
\bea
             P_{\rm cap}(s,r) \;=\;  p(s) -  b_{-2}(s) r^{-1} + \sum_{k = 1}^\infty b_{k-1}(s)\, r^k/k\,,
\label{PcapSeries}
\eea
where the quantity $p(s)$ (independent of $r$) appears as the integration ``constant''. It turns out that if the 
tube caps $\Sigma_{\rm cap}(s,r)$ are orthogonal to the worldline ${\cal C}$, as they are for the Dirac-type tube being 
used here, then this integration constant can be directly 
identified with the momentum of the particle (the situation for a worldtube with null-cone caps is
covered in Section~\ref{id_of_p}).
Differentiating (\ref{PcapSeries}) with respect to $s$ finally gives 
\be
    \frac{\partial P_{\rm cap}(s,r)}{\partial s} \;=\;  \dot p(s) - \dot b_{-2}(s) r^{-1} 
                                                                  + \sum_{k = 1}^\infty  \dot b_{k-1}(s)\, r^k/k\,.
\label{secondexp}
\ee
We have thus found a {\em second} expression for the quantity $\partial_s P_{\rm cap}(s,r)$, in addition to Dirac's balance 
relation (\ref{DiracBalance}).
Subtracting (\ref{secondexp}) from (\ref{DiracBalance}) then gives
\be
      0 \;=\; \left(a_{-1} + \dot b_{-2}\right) r^{-1} + (a_0 - \dot p) + 
                      \sum_{k = 1}^\infty  \left(a_k - \dot b_{k-1}/k \right)\, r^k \,.
\label{zero_series}
\ee
Since $r$ is arbitrary, vanishing of this series entails that all coefficients in the series are 
zero. The $r^0$-term gives an equation to be satisfied by $p(s)$, whereas the terms involving $r$ vanish 
identically\footnote{For $k\ne 0$, the coefficients $a_k(s)$ are therefore perfect differentials. Having noticed this
before obtaining (\ref{zero_series}), the author wrote a Mathematica package for symbolic integration of 
Frenet frame expressions such as (\ref{a1coeff}). This package is available as part of \cite{Norton:08}.}. Using the 
Mathematica code \cite{Norton:08}, one can check this is so to at least $\Or\!\left(r^5\right)$. We remark that to explicitly 
verify that the coefficient of $r^k$ is identically zero requires, for $k \ge 2$, that on ${\cal C}$ the background field $F$ 
must satisfy the source-free Maxwell equations and also derivatives of these equations (up to a derivative order 
that increases with $k$). The vanishing of the $r^0$-term in (\ref{zero_series}) gives the momentum balance equation, $\dot p = a_0$.
For the Lorentz--Dirac example, the coefficient $a_0$ is given by (\ref{a0coeff}) and this equation becomes   
\be
              \dot p(s) \;=\;  q F\cdot N_0 +  \frac{q^2}{6 \pi\epsilon_0 c} 
                                     \left(\tensor{\dot \kappa}{_1} \N{_1} + \kap{_1}\kap{_2}\N{_2} \right)\,.
\label{mybalance}
\ee
This is an exact statement of momentum balance for the case in which 
the asymptotic field of the charged particle is equal to the retarded 
field of an electric monopole. 

Note that $p(s)$ is still unspecified. 
To obtain an equation of motion from (\ref{mybalance}), one must specify how  
the particle momentum is to be defined in terms of the geometry of the worldline and the internal degrees of freedom
of the radiating system or particle. The Lorentz--Dirac equation is obtained by assuming (as does Dirac) that 
the particle has no internal degrees of freedom and that $p(s)$ takes the simplest possible form,
\be
                                      p(s) \;=\; cm \N{_0_\!}(s)\,.
\label{pdef}
\ee
Then (\ref{mybalance}) becomes the Lorentz--Dirac equation, the equivalence of (\ref{mybalance}) and (\ref{dirac_eom1}) being 
easily established using the Frenet--Serret equations (\ref{dot_N_0})--(\ref{dot_N_3}).  

There are several logical gaps in the above derivation of the Lorentz--Dirac equation, but these gaps are not related to 
evaluating the self-force. If one knows the asymptotic electromagnetic field of the extended particle (even if this field 
is defined so as to depend on internal degrees of freedom of the particle) then the tube method can be used to derive an 
exact statement of momentum balance analogous to (\ref{mybalance}), that defines the self-force. 
The gaps only become evident when attempting to derive an
equation of motion from this balance equation. For example, definitions of the momentum and center-of-mass worldline 
of an extended particle have been given by Dixon in \cite{Dixon:70a}--\cite{Dixon:74}. But our $p(s)$ is defined as the 
the $r$-independent term (the integration constant) in the series (\ref{PcapSeries}) for $P_{\rm cap}(s,r)$.
Is our $p(s)$ consistent with Dixon's definition of momentum? Does the worldline ${\cal C}$ 
have to be chosen to coincide with the center-of-mass worldline? We leave these questions for future investigation.
For more on the problem of deriving equations motion for radiating extended charged particles, see \cite{Harte:06}.

\subsection{Classical mass renormalization}
\label{cmr}

The classical mass renormalization procedure
can be understood as a short-cut that amounts to assuming an
$\Or(r)$ approximation to our expression (\ref{PcapSeries}). To see this, we reproduce here the 
argument used in \cite{Dirac:38} to get from Dirac's balance relation (\ref{DiracBalance}) to the result (\ref{mybalance}). 
 
The traditional interpretation of the quantity $P_{\rm cap}(s,r)$ has been that in the limit $r\rightarrow 0$
this represents a ``bare momentum'' of the particle. Assume now that as $r\rightarrow 0$ the bare momentum 
takes an almost familiar form, being the product of the 4-velocity of the particle with a bare mass function $\mB(r)$.
Thus, (compare equation (20) in \cite{Dirac:38}) let 
\be
                     P_{\rm cap}(s,r) \;=\; c\mB(r) \N{_0_\!}(s) + \Or(r)\,.  
\label{Pguess}  
\ee    
One now chooses the bare mass to be negative and diverging as $r\rightarrow 0$ in exactly the manner required so as to cancel
the problematic $r^{-1}$ term appearing on the right-hand side of (\ref{DiracBalance}). Moreover, the difference between these 
divergent terms must account for the physically observable mass of the particle. Thus, (compare equation (21) in \cite{Dirac:38})
one postulates that
\be
                c \mB(r) \;=\; - \frac{q^2}{8 \pi \epsilon_0 c} \,\frac{1}{r} + c m\,,    
\label{mB}     
\ee
where $m$ is now referred to as the ``renormalized'' mass of the particle. Substituting (\ref{mB}) into (\ref{Pguess}) gives
\be
              P_{\rm cap}(s,r) \;=\; c m \N{_0} - \frac{q^2}{8 \pi \epsilon_0 c} \,\frac{\N{_0}}{r} + \Or(r)\,, 
\label{Pcap_approx}     
\ee
which is recognised as being the $\Or(r)$ approximation to (\ref{PcapSeries}), for a particle momentum
given by (\ref{pdef}). So an $\Or(r)$ approximation to the Lorentz--Dirac equation is obtained by following the steps 
previously given. The remaining $r$-dependent terms are then eliminated from the equation of motion by taking $r\rightarrow 0$, 
in which case (\ref{mB}) becomes an infinite mass renormalisation.
 
The renormalization procedure evidently depends on being able to antidifferentiate with respect to $s$ any singular terms that appear on 
the right hand side of (\ref{int1}) so that these can be absorbed into a redefined momentum on the left hand side of 
(\ref{int1}). In 1944, Bhabha and Harish-Chandra \cite{Bhabha:44} showed that for point particles possessing any multipole 
moments whatsoever, this can always be done. Using our present notation, what they found is that (compare equation (19) in 
\cite{Bhabha:44}),
\be  
           \frac{\partial}{\partial r} \int_{\Omega(s,r)} T\cdot \nu_{\rm tube} \,d\Omega
         + \frac{\partial}{\partial s} \int_{\Omega(s,r)} T\cdot \nu_{\rm cap} \,d\Omega \,=\,  0 \,,
\label{commute_id}
\ee  
which is the relation $(\partial_r\partial_s - \partial_s\partial_r)P_{\rm cap}(s,r)=0$ expressed in terms of (\ref{PcapDr}) and (\ref{int1}).
They go on to observe that if $T$ can be expanded as a power series in $r$, then (\ref{commute_id}) shows that, with the exception of the
$r^0$-term, all coefficients in the series for (\ref{int1}) (which they call ``the inflow'') are perfect differentials. 
In a later paper \cite{Bhabha:46}, the authors refer to this result as their inflow theorem. 
In particular, in \cite{Bhabha:44} they say 
``the singular terms being perfect differentials can always be compensated 
by the addition of suitable terms to $P_{\rm cap}(s,r)$''\footnote{The original text refers to a quantity $A(\tau)$  
that corresponds to our $P_{\rm cap}(s,r)$ in (\ref{int1}).}. These two papers are both concerned with point particle theory 
(the phrase ``point particles'' even appears in both titles) and
there is no indication that the authors were aware of the full significance of their inflow theorem:
If all of the perfect differentials, not just the singular terms, are
``compensated'' then all dependence on $r$ vanishes, exactly as it did in equation (\ref{zero_series}). There is then no reason to take 
$r \rightarrow 0$ and the resulting balance equation (the counterpart of (\ref{mybalance})) is exact for finite radius worldtubes.   
Reference \cite{Bhabha:44} has received 
little attention (an exception is \cite{vanWeert:74} page 361, but again, in the context of renormalization for point particles) and the 
fact that the inflow theorem underlies a finite radius tube method seems to have been overlooked.

\subsection{Identification of the particle momentum}
\label{id_of_p}

It was remarked earlier that being able to identify the integration constant in (\ref{PcapSeries}) 
as the particle momentum depended on having chosen to use a worldtube with caps that are orthogonal to 
the worldline. To see how a result that is independent of tube cap geometry can come about, one can do the
calculations using a variety of worldtubes, with differing tube cap geometries, then demand
that the result not depend on any particular choice of worldtube. Thus, in Section~\ref{tube_coordinates} 
we define a 1-parameter family of worldtubes, with parameter
$\varepsilon \in [-1,1]$. The tube caps vary with $\varepsilon$, interpolating between 
forward null cones for $\varepsilon = 1$ and backward null cones for $\varepsilon = -1$.
The Dirac-type worldtube, with caps orthogonal to ${\cal C}$, corresponds to $\varepsilon = 0$.

In this more general setting, the integration constant in (\ref{PcapSeries}) depends on
$\varepsilon$, and thus $p(s)$ in (\ref{PcapSeries}) is to be replaced by $p_\varepsilon(s)$.
In place of (\ref{mybalance}) one then finds that \cite{Norton:08},
\be
         \dot p_\varepsilon(s) \;=\;  q F\cdot N_0 +  \frac{q^2}{6 \pi\epsilon_0 c} 
                                     \left( -\varepsilon\kap{_1}^2 \N{_0}
            + \left(1 -\varepsilon\right) \left( \tensor{\dot \kappa}{_1} \N{_1} + \kap{_1}\kap{_2}\N{_2} \right)\right).
\ee
An equation that is independent of $\varepsilon$ is obtained by integrating the $\varepsilon$-dependent
terms and absorbing them into the definition of the integration constant. Thus, by setting
\be
          p_\varepsilon(s) \;=\; p(s) - \frac{q^2}{6 \pi\epsilon_0 c} \, \varepsilon \kap{_1}\N{_1},
\ee  
one recovers the relation (\ref{mybalance}). 

\subsection{A sanity check --- the weak energy condition}
\label{sanity_ck}

In Section \ref{cmr} we saw that 
if the classical mass renormalization procedure is fully implemented using Bhabha and Harish-Chandra's inflow theorem, 
then one recovers (in a somewhat awkward manner) the finite radius tube method of Section \ref{tube_method}. 
Taking $r \rightarrow 0$ is then quite pointless and 
only serves to make the theory mathematically ill-defined and physically nonsensical.
In particular, the bizarre notion of negative bare mass (equations (\ref{mB})--(\ref{Pcap_approx})) arises because 
the worldtube is too small to fit the particle within it. Using the weak energy condition, this can be turned around to determine
a minimum possible size for the extended particle or system by finding the smallest worldtube that could enclose a system
with the same prescribed multipole properties.

If at every point the stress-energy tensor obeys the inequality $\tensor{T}{_\alpha_\beta} \tensor{u}{^\alpha} \tensor{u}{^\beta} \ge 0$ for any 
timelike vector $u$, then $T$ is said to satisfy the weak energy condition \cite{HawkingEllis:73}. This is equivalent to the energy 
density being non-negative for any observer. Assuming the weak energy condition for $T$, one has from (\ref{Pcap}) that
\be
          - \N{_0}\!  \cdot P_{\rm cap}(s,r) \;=\; \frac{1}{c}\int_{r^\prime = 0}^r  \int_{\Omega(s,r^\prime)}
                                          \tensor{T}{_\alpha_\beta} \N{_0^\alpha} \N{_0^\beta} \,\rmd\Omega\,\rmd r^\prime \;\ge\; 0\,,  
\ee
where we have used a Dirac-type worldtube, for which (\ref{nu_cap}) gives $\nu_{\rm cap} = -\N{_0}$. Substituting expression (\ref{PcapSeries}) for 
$P_{\rm cap}$, which we saw in Section~\ref{id_of_p} is specifically for a Dirac-type worldtube, one has
\be
                        - \N{_0}\cdot p +  \N{_0}\cdot b_{-2}(s) \,r^{-1} + \Or(r) \;\ge\; 0\,.
\label{gen_ineq}
\ee
If the asymptotic field of the particle is that of an electric monopole, then $b_{-2}(s)$ is given by (\ref{b_2}). In this case, after 
dropping $\Or(r)$ terms and rearranging, (\ref{gen_ineq}) can be written as
\be
                   r \,\ge\, \frac{1}{8\pi\epsilon_0} \,\frac{q^2}{m c^2}\, \frac{cm}{(- \N{_0}\cdot p)}\,. 
\label{sanity}
\ee
It is interesting to see what this result says about classical models for the electron. In the Lorentz--Dirac model the momentum is assumed
to be tangent to the worldline of the particle and thus, $cm/(-\N{_0}\cdot p) = 1$. Inequality (\ref{sanity}) then becomes $r \ge r_{\rm e}/2$, where
$r_{\rm e}$ is the so-called classical electron radius \cite{Jackson:98},
\be
                 r_{\rm e} \,\equiv \, \frac{1}{4\pi\epsilon_0} \,\frac{e^2}{m_{\rm e}c^2} 
                            \,\approx\, 2.8 \times 10^{-15}\,{\rm m}\,,             
\ee
where $q = -e$ is the electron charge and $m = m_{\rm e}$ is the electron mass. However, it is known experimentally \cite{Kopp:95} that the 
minimum tube radius for an electron is at least $10^3$ times smaller than $r_{\rm e}/2$. Not surprisingly, it follows that the Lorentz--Dirac classical 
electron model would be completely inadequate for describing the physics in such experiments. On the other hand, our result (\ref{sanity}) can not be 
used to rule out all classical models of 
a {\it spinning} electron. In a classical spin model the worldline is typically a spacetime helix. The particle momentum is not tangent to the worldline, 
but instead, is directed along the helix axis. The factor $cm/(-\N{_0}\cdot p)$ appearing in (\ref{sanity}) can therefore be arbitrarily small if the 
circular motion of the charge centre is close to the speed of 
light\footnote{Calculating in the momentum rest frame, one has $p = (cm,{\bf 0})$ and $\N{_0}=\gamma (1,{\bf v}/c)$ where 
$\gamma = (1 - (v/c)^2)^{-1/2}$ and ${\bf v}$ is the 3-velocity of the circular motion. Then, $cm/(-\N{_0}\cdot p) = \gamma^{-1}$.}. 
Such models arise naturally in 2nd and higher order Lagrangian mechanics (for example, see \cite{Nesterenko:96, Arreaga:01}).

\section{Geometry of the worldtube}
\label{tube_coordinates}

At least two obvious choices present themselves for the worldtube. One is a worldtube with 
caps orthogonal to ${\cal C}$, as used by Dirac \cite{Dirac:38}, the other is a worldtube with 
caps that are parts of forward null-cones based on ${\cal C}$, sometimes called a Bhabha tube and 
used, for example, in \cite{Bhabha:44}, \cite{Rowe:75} and \cite{Poisson:99}. 
In either case, $\Sigma_{\rm tube}(s,r)$ and $\Sigma_{\rm cap}(s,r)$ can be defined by equations
(\ref{tube_defn}) and (\ref{cap_defn}) as coordinate surfaces in a system of spherical polar tube 
coordinates $\{s, r, \vartheta, \varphi\}$ centred on the worldline ${\cal C}$, with the 
angles $\{\vartheta, \varphi\}$ defined with respect to the Frenet frame of ${\cal C}$. 

\subsection{Spherical polar tube coordinates}

\subsubsection{The Frenet frame of ${\cal C}$}
\label{frenet_frame}

Recall that the world-line ${\cal C}$ has the equation $x = z(s)$, where $x^\alpha$ are Minkowski coordinates 
and $s$ is proper distance along ${\cal C}$. Let the Frenet frame vectors of ${\cal C}$ be 
\be
                \N{_i}(s) \,=\, \N{_i^\beta}(s) \,\tensor{\bf e}{_\beta}\,,
\label{N_defn}
\ee
where $\tensor{\bf e}{_\beta}$ are Minkowski basis vectors, 
$\tensor{\bf e}{_\alpha}\cdot \tensor{\bf e}{_\beta} = \tensor{\eta}{_\alpha_\beta}$. The defining relations for the 
Frenet frame are that $\N{_0} = \dot z = \rmd z/\rmd s$ is the future-directed unit tangent 
to ${\cal C}$; that the frame $\N{_i}$ is orthonormal and has the same orientation as the Minkowski frame 
$\tensor{\bf e}{_\beta}$; and that the frame $\N{_i}$ satisfies the Frenet--Serret equations
\bea
              \tensor{\dot N}{_0} &= \kap{_1} \N{_1}\,,
\label{dot_N_0}
\\
              \tensor{\dot N}{_1} &= \kap{_1} \N{_0} 
                                    + \kap{_2} \N{_2} \,,
\label{dot_N_1}
\\
              \tensor{\dot N}{_2} &= - \kap{_2} \N{_1} 
                                      + \kap{_3} \N{_3} \,,
\label{dot_N_2}
\\
              \tensor{\dot N}{_3} &= - \kap{_3} \N{_2}\,.
\label{dot_N_3}
\eea
The scalar fields $\kap{_a_\!}(s)$ are the 
Frenet--Serret curvatures of ${\cal C}$. Differentiation with respect to proper distance $s$ is  
indicated by an overdot. The orthogonality conditions for the Frenet frame are 
$\N{_i} \cdot \N{_j} = \tensor{\eta}{_i_j}$, where $\eta = \diag(-1,1,1,1)$, or in terms of components,
$ 
        \N{_i^\alpha} \N{_j^\beta} \tensor{\eta}{_\alpha_\beta}=\tensor{\eta}{_i_j}.
$
The matrix inverse of $\N{_i^\alpha}$ is therefore given by its Minkowski transpose,
$
           \N{^k_\alpha} \,=\,  \tensor{\eta}{^k^j} \N{_j^\beta} \tensor{\eta}{_\alpha_\beta}
$. 
Finally, one has the following relation between the Minkowski basis covectors and the Frenet coframe,
\be
               \tensor{\bf e}{^\alpha} \,=\, \N{^i} \N{_i^\alpha}\,,       
\label{co_reln}
\ee
where $\tensor{\bf e}{^\alpha} = \tensor{\eta}{^\alpha^\beta} \tensor{\bf e}{_\beta}$,  
and $\N{^i} = \tensor{\eta}{^i^j} \N{_j}$. 

\subsubsection{Time-symmetric tube coordinates}
\label{symmetric_tube_coords}

Time-symmetric tube coordinates $\{s, r, \vartheta, \varphi\}$, centred on the world-line ${\cal C}$ are defined by the 
coordinate transformation
\be
                        x  \,=\, z(s) + r \,n(s,\,\vartheta,\,\varphi)\,.
\label{xpos}
\ee
Here $x = x^\alpha\tensor{\bf e}{_\alpha}$ and $z(s)= z^\alpha(s)\tensor{\bf e}{_\alpha}$ 
are Minkowski position vectors, and $n$ is a unit vector orthogonal to ${\cal C}$, parametrized in terms of polar angles as
\be
           n \,=\, \N{_1_\!}(s) \sin\vartheta \cos\varphi
                 + \N{_2_\!}(s) \sin\vartheta \sin\varphi
                 + \N{_3_\!}(s) \cos\vartheta\,. 
\label{n_defn}
\ee  
The surfaces $s={\rm const.}$ are 3-planes orthogonal to ${\cal C}$ at $x=z(s)$.
Within these 3-planes the coordinates $\{r, \vartheta, \varphi\}$ are standard spherical polar coordinates,
with the polar angles measured with respect to the directions defined by the Frenet frame at $x=z(s)$. 
We call these coordinates ``time-symmetric'' rather than ``instantaneous rest-frame'' or similar because
the Frenet frame is Fermi-propagated (non-rotating) along ${\cal C}$ only if $\kap{_2}=\kap{_3}=0$. Note also, that 
implicit in (\ref{xpos}) is the identification $\N{_i}(s,r,\vartheta,\varphi)\equiv \N{_i}(s)$,
that is, the frame at $x = x(s,r,\vartheta,\varphi)$ is obtained by parallel transport of $\N{_i}(s)$ from the point 
$x = z(s)$. Thus, the Frenet frame, originally defined only on ${\cal C}$, is henceforth considered to be a frame 
field on Minkowski 
spacetime\footnote{This frame field is only needed in the vicinity of ${\cal C}$,
where the tube coordinate metric  $g$ is non-singular. For the coordinate transformation (\ref{xpos_eps}), which includes (\ref{xpos})
for $\varepsilon = 0$, one has  
$\sqrt{-\det(g)} = r^2\sin\vartheta \,(1 + r \kap{_1}(1 - \varepsilon^2) \sin\vartheta\cos\varphi )$ near ${\cal C}$,
so $g$ is non-singular if $|r\kap{_1}| < (1-\varepsilon^2)^{-1}$.}.

\begin{figure}[h!tb]
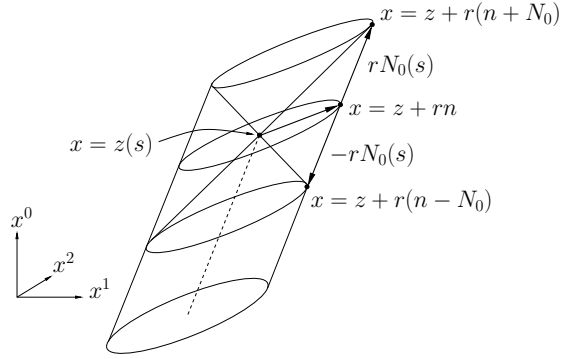

        \centering
        \resizebox{7cm}{!}{\input norton_figure2.tex}
        \caption{The tube caps are $s={\rm const.}$ surfaces in one of the tube coordinate systems defined by 
        the transformation $x = z + r(n + \varepsilon N_0)$. Shown here are the
        null caps for $\varepsilon = \pm 1$, and the Dirac-type cap for $\varepsilon = 0$ 
        (the $r = {\rm const.}$ worldtubes in this figure coincide for different $\varepsilon$ 
        only because the figure has been drawn for a source in linear motion).}   
        \label{capfig}
\end{figure}

\subsubsection{Retarded (advanced) tube coordinates}
\label{ret_tube_coords}

Retarded (advanced) tube coordinates are obtained by taking the $s={\rm const.}$ surfaces to be the
forward (backward) null-cones based on ${\cal C}$.
The coordinates $\{s, r, \vartheta, \varphi\}$ are defined by the transformation
\be
                   x  \,=\, z(s) + r \left( n(s,\,\vartheta,\,\varphi) + \varepsilon \N{_0_\!}(s) \right),
\label{xpos_eps}
\ee
where $\varepsilon=1$ gives the retarded coordinate system and $\varepsilon=-1$ gives the advanced system,
as shown in Figure~2.
Transformation (\ref{xpos_eps}) can also be interpreted as defining a 
continuous 1-parameter family of tube coordinate systems. 
Since $\varepsilon=0$ corresponds to the time-symmetric tube coordinates (\ref{xpos}), we may deal with all interesting 
cases at once by carrying $\varepsilon\in [-1,1]$ as a parameter through our calculations. 

\subsection{Gauss' formula for the worldtube}
\label{gauss_formula}

We require Gauss's formula specialised to integrating over the tube volume (\ref{volume_defn}). 
Writing the boundary of $V(s,r)$ as 
$\partial V(s,r) = \Sigma_{\rm cap}(s,r) - \Sigma_{\rm cap}(0,r) + \Sigma_{\rm tube}(s,r)$ one has that for
any $C^1$ vector field $X$, 
\bea
\fl   \int_{V(s,r)} \nabla\cdot X\,dV(s,r) &=& 
                                                  \int_{\Sigma_{\rm cap}(s,r)}X \cdot d\Sigma_{\rm cap}(s,r)
                                                - \int_{\Sigma_{\rm cap}(0,r)}X \cdot d\Sigma_{\rm cap}(0,r)
\nonumber \\ && {}
                                                + \int_{\Sigma_{\rm tube}(s,r)}X \cdot d\Sigma_{\rm tube}(s,r)\,.
\label{gauss}
\eea 
In Sections \ref{integration_over_hypersurface}--\ref{elt_for_worldtube} it is shown that these surface integrals 
can be evaluated as 
\bea
      \int_{\Sigma_{\rm cap}(s,r)}X \cdot d\Sigma_{\rm cap}(s,r) &=&  \int_{r^\prime = 0}^r  \int_{\Omega(s,r^\prime)}
                                                   X\cdot \nu_{\rm cap}\,\rmd\Omega\,\rmd r^\prime\,,
\label{Xcap}
\\                   
      \int_{\Sigma_{\rm tube}(s,r)}X \cdot d\Sigma_{\rm tube}(s,r) &=&   \int_{s^\prime = 0}^s  \int_{\Omega(s^\prime,r)} 
                                                   X\cdot \nu_{\rm tube}\,\rmd \Omega\,\rmd s^\prime\,.    
\label{Xtube}
\eea
Here $\rmd \Omega$ is the volume element of the 2-sphere $\Omega(s,r)$. 
It follows that the orientations of $\Sigma_{\rm tube}(s,r)$ and 
the final ($s>0$) tube cap $\Sigma_{\rm cap}(s,r)$, are given by the directions of the normals $\nu_{\rm tube}$ 
and $\nu_{\rm cap}$ 
respectively, and for (\ref{gauss}) to hold these are to be chosen so that $X \cdot \nu$ is positive if $X$ is a vector 
that points out of $V(s,r)$
(see for example Section 2.8 in \cite{HawkingEllis:73}, or Appendix B.2 in \cite{Wald:84}). The initial ($s=0$) 
tube cap is then reverse-oriented with respect to $\nu_{\rm cap}$ 
since $X \cdot \nu_{\rm cap}$ is negative for outward pointing vectors on $\Sigma_{\rm cap}(0,r)$.
This is in accordance with the sign of the corresponding term in (\ref{gauss}). 
The normals are given by (\ref{nu_cap}) and (\ref{nu_tube}). 

\subsubsection{Energy-momentum conservation}
\label{conservation}

Let $T = \tensor{T}{^\alpha^\beta}\tensor{\bf e}{_\alpha}\otimes\tensor{\bf e}{_\beta}$ be the conserved stress-energy
tensor (\ref{Tdefn}). The four vector fields (for $\alpha = 0,\ldots,3$) defined by 
$\tensor{X}{^{(\alpha)}} = \tensor{\bf e}{^\alpha}\cdot T = \tensor{T}{^\alpha^\beta}\tensor{\bf e}{_\beta}$ are then divergence
free, $\nabla \cdot \tensor{X}{^{(\alpha)}} = 0$. Applying Gauss's formula (\ref{gauss}) to each of these vector fields
gives the four conservation equations,
\be
            0 \,=\, P^{\,\alpha}_{\rm cap}(s,r) - P^{\,\alpha}_{\rm cap}(0,r) + P^{\,\alpha}_{\rm tube}(s,r)\,,
\label{plaws}
\ee 
where
\bea
      P^{\,\alpha}_{\rm cap}(s,r) &=& \frac{1}{c}\, \int_{\Sigma_{\rm cap}(s,r)} \tensor{X}{^{(\alpha)}} \cdot d\Sigma_{\rm cap}(s,r)\,,
\\
      P^{\,\alpha}_{\rm tube}(s,r) &=& \frac{1}{c}\, \int_{\Sigma_{\rm tube}(s,r)} \tensor{X}{^{(\alpha)}} \cdot d\Sigma_{\rm tube}(s,r)\,.
\eea
The momentum flux 4-vectors are defined by $P_{\rm cap} = \tensor{\bf e}{_\alpha} P^{\,\alpha}_{\rm cap}$ and 
$P_{\rm tube} = \tensor{\bf e}{_\alpha} P^{\,\alpha}_{\rm tube}$. Equations (\ref{Pcap})--(\ref{Ptube}) then follow from
(\ref{Xcap})--(\ref{Xtube}), and the energy-momentum conservation law (\ref{plaw}) follows from (\ref{plaws}). 
                    
\subsubsection{The directed hypersurface element $\rmd\Sigma$}
\label{integration_over_hypersurface}

Suppose we are given a hypersurface $\Sigma$ defined in terms of spherical polar tube coordinates. That is,  
$\Sigma$ is parametrized by coordinates $y^A$, $A=1,2,3$ and has equations of the 
form $(s,r,\vartheta,\varphi)= (f^\alpha(y^A))$. We shall derive here the 
Frenet frame expression for the directed surface element of $\Sigma$.

In Minkowski coordinates the directed hypersurface element is $\rmd \Sigma = \tensor{\bf e}{^{\mu\,}}\rmd \Sigma_\mu$,
where the 3-forms $\rmd \Sigma_\mu$ are given by 
\be
        \rmd \Sigma_\mu \,=\, \frac{1}{3!}\, \tensor{\varepsilon}{_\mu_\alpha_\beta_\gamma} 
                           \, \rmd x^\alpha \wedge \rmd x^\beta \wedge \rmd x^\gamma\,.
\label{sigmu}
\ee
Here $\tensor{\varepsilon}{_\mu_\alpha_\beta_\gamma}$ is the fully antisymmetric Levi-Civita tensor
($\tensor{\varepsilon}{_0_1_2_3}=1$ in Minkowski coordinates). For $\Sigma$ parametrized 
by coordinates $y^A$, the 1-forms $\rmd x^\alpha$ in (\ref{sigmu}) can be evaluated as 
\bea
          \rmd x^\alpha \,=\, \frac{\partial}{\partial y^A}\left(\tensor{\bf e}{^\alpha} \cdot x\right) \rmd y^A   
                    \,=\, \tensor{\bf e}{^\alpha} \cdot \frac{\partial x}{\partial y^A} \, \rmd y^A\,. 
\label{dxdy}      
\eea 
The vectors $\partial x/\partial y^A$ that appear here are to be calculated by differentiating (\ref{xpos_eps}),
so are naturally obtained as linear combinations of the Frenet frame vectors. Let these linear combinations be
\be
             \frac{\partial x}{\partial y^A} \,=\, \N{_j} \tensor{X}{^j_A}\,.  
\label{XjA}   
\ee
Using (\ref{co_reln}), equation (\ref{dxdy}) now becomes 
\be
       \rmd x^\alpha  \,=\, \tensor{\bf e}{^\alpha} \cdot \N{_j} \tensor{X}{^j_A}  \rmd y^A
                  \,=\, \N{_j^\alpha} \tensor{X}{^j_A} \rmd y^A\,.
\label{dx}
\ee 
Again using (\ref{co_reln}), one has $\rmd \Sigma = \N{^i} \N{_i^\mu} \rmd \Sigma_\mu$.
Substituting (\ref{dx}) into (\ref{sigmu}) then gives
$$
          \rmd \Sigma  \,=\, 
                   \frac{1}{3!}\, \N{^i} \,\tensor{\varepsilon}{_\mu_\alpha_\beta_\gamma} \,
                                  \N{_i^\mu} \N{_a^\alpha} \N{_b^\beta} \N{_c^\gamma} 
                                  \tensor{X}{^a_A} \tensor{X}{^b_B} \tensor{X}{^c_C} \,
                                  \tensor{\rmd y}{^A} \wedge \tensor{\rmd y}{^B} \wedge \tensor{\rmd y}{^C}\,, 
$$
and because the Frenet frame is orthonormal and has the same orientation as the Minkowski frame,
\be
             \tensor{\varepsilon}{_\mu_\alpha_\beta_\gamma} \,
                                  \N{_i^\mu} \N{_a^\alpha} \N{_b^\beta} \N{_c^\gamma} 
           \,=\, \tensor{\varepsilon}{_i_a_b_c}\,. 
\ee
The hypersurface element therefore simplifies to 
\be 
                \rmd \Sigma  \,=\,    \N{^i} \,\tensor{\varepsilon}{_i_a_b_c}\,
                                      \tensor{X}{^a_1} \tensor{X}{^b_2} \tensor{X}{^c_3} \,
                                      \tensor{\rmd y}{^1} \wedge \tensor{\rmd y}{^2} \wedge  \tensor{\rmd y}{^3}\,, 
\label{sig}
\ee 
which is the required Frenet frame expression. 

\subsubsection{Directed surface element for a tube cap}

The tube cap $\Sigma_{\rm cap}(s,r)$ is defined by (\ref{cap_defn}) as part of the $s={\rm const.}$ coordinate surface. 
For $\varepsilon=0$ the caps are orthogonal to ${\cal C}$, 
whereas for $\varepsilon=\pm1$ they are null-cones based on ${\cal C}$. As parameters for the tube cap, one may take 
$\{y^A\} = \{r,\vartheta,\varphi\}$. The derivatives $\partial x/\partial y^A$ are found using (\ref{xpos_eps}),
\bea
         \partial x/\partial r &= n + \varepsilon \N{_0}\,,
\\
         \partial x/\partial \vartheta  &= r \left(  \N{_1} \cos\vartheta \cos\varphi
                                                            + \N{_2} \cos\vartheta \sin\varphi
                                                            - \N{_3} \sin\vartheta \right),
\label{dxdtheta}
\\
         \partial x/\partial \varphi  &= r \left( - \N{_1} \sin\vartheta \sin\varphi
                                                             + \N{_2} \sin\vartheta \cos\varphi \right).
\label{dxdphi}
\eea
From these expressions one can read off the coefficients $\tensor{X}{^j_A}$, as defined by (\ref{XjA}).  
The directed hypersurface element (\ref{sig}) is found to be
\be
          \rmd \Sigma \,=\, \nu_{\rm cap}\, \rmd \Omega \wedge \rmd r  \,,  
\ee 
where $\rmd \Omega =  r^2 \sin\vartheta\, \rmd\vartheta \wedge \rmd\varphi$, and the normal is
\be
                \nu_{\rm cap} \,=\, - (\N{_0} + \varepsilon n)\,,
\label{nu_cap}
\ee
where the sign of $\nu_{\rm cap}$ has been chosen as described in Section (\ref{gauss_formula}).

\subsubsection{Directed surface element for the worldtube}
\label{elt_for_worldtube}

The worldtube $\Sigma_{\rm tube}(s,r)$ is defined by (\ref{tube_defn}) as part of the $r={\rm const.}$ coordinate surface. 
As parameters for the worldtube, one may 
take $\{y^A\} = \{s,\vartheta,\varphi\}$. The derivatives $\partial x/\partial y^A$ that are needed 
to evaluate the surface element are (\ref{dxdtheta})--(\ref{dxdphi}) together with
\bea
         \partial x/\partial s \,=\, \N{_0} + r \left( \dot n + \varepsilon \kap{_1} \N{_1} \right),
\label{dxds}
\eea
where 
$$
       \dot n \,=\,   \left( \kap{_1} \N{_0} + \kap{_2} \N{_2} \right) \sin\vartheta \cos\varphi 
                    + \left( - \kap{_2} \N{_1} + \kap{_3} \N{_3} \right) \sin\vartheta \sin\varphi 
                    - \kap{_3} \N{_2}  \cos\vartheta\,.
$$
For the coordinate ordering specified by $\{y^A\} = \{s,\vartheta,\varphi\}$, one finds that (\ref{sig}) gives
\be
             \rmd \Sigma \,=\, - \nu_{\rm tube}\, \rmd \Omega  \wedge \rmd s \,.  
\label{tube_elt}
\ee 
where the normal
\be
            \nu_{\rm tube} \,=\, n + r \kap{_1} \sin\vartheta \cos\varphi \left(n + \varepsilon \N{_0} \right),
\label{nu_tube}
\ee
has its sign chosen as described in Section (\ref{gauss_formula}). 



\section{The stress-energy tensor}
\label{stress_energy_tensor}

For $r>r_0$, equation (\ref{Tdefn}) defines $T$ to be the stress-energy tensor of an electromagnetic field. We take this field 
to be the sum of a background field $F$ and a prescribed retarded field $F_\ret$ for the particle or radiating source 
that we are interested in,
\be
                         F_{\rm tot} \;=\; F + F_\ret\,.
\label{Ftot}
\ee 
The background field $F$ will include any applied field, any fields generated by other moving charged particles, and also any
radiation from the source that has been reflected back to the source position\footnote{This last 
contribution to the background field accounts for changes in spontaneous emission rates and effects thereof, such as bandgaps in photonic crystals.} 
by boundaries or material media. Thus, $F$ will 
appear in the calculations as an unspecified solution of the Maxwell equations, known to be source-free throughout the 
the spacetime region occupied by the particle or system of interest.

The stress-energy tensor corresponding to $F_{\rm tot}$ is $T_{\rm e.m.} = \tensor*{T}{_{\rm e.m.}^i^j}\,\N{_i}\otimes\N{_j}$, where the 
frame components are
\be
         \tensor*{T}{_{\rm e.m.}^i^j} \;=\; \frac{1}{\mu_0} 
                  \left( \tensor*{F}{_{\rm tot}^i^a} \tensor*{F}{_{\rm tot}^j^b} \tensor{\eta}{_a_b} 
      - \frac{1}{4} \,\tensor{\eta}{^i^j} \tensor*{F}{_{\rm tot}^a^c} \tensor*{F}{_{\rm tot}^b^d} \tensor{\eta}{_a_b} \tensor{\eta}{_c_d} 
                 \right). 
\label{Tem}
\ee
For the Lorentz--Dirac charged particle, the field of the radiating source is taken to be the retarded {\LW} field.

\subsection{The \LW\ field}
\label{LW_field}

The retarded \LW\ potential for the electromagnetic field of a point
charge $q$ with worldline ${\cal C}$ is
\be
          A(x) \,=\, \frac{q}{4\pi\epsilon_0 c} \,\frac{\dot z_\ret(x)}{R(x)} \,,
\label{phi_ret}
\ee
where the vector and scalar fields $\dot z_\ret(x)$ and $R(x)$ are defined as follows. Let the retarded proper distance
$s_\ret(x)$ be the scalar field that gives the worldline parameter value of the source-point on ${\cal C}$ for a retarded 
field at $x$. Thus, $s_\ret(x)$ is defined so that $x$ lies on the forward null-cone based at $z(s_\ret(x))$, and
\be
                                 \zeta(x) \;=\; x - z(s_\ret(x))\,,
\label{zeta}
\ee
%
%
is therefore a future directed null vector field. Then $\dot z_\ret(x) = \dot z(s_\ret(x))$ is 
defined as the unit tangent to ${\cal C}$ parallel propagated from the retarded source point $z_\ret(x)=z(s_\ret(x))$. Similarly,
one defines $\ddot z_\ret(x) = \ddot z(s_\ret(x))$. The scalar field $R$ is the radial distance of $x$ from the 
source, as measured in the instantaneous rest-frame of the source at the source-point $z_\ret$,
\be
                                 R \,=\,  - \dot z_\ret \cdot \zeta\,.  
\label{def_R}        
\ee 
The scalar and 3-vector potentials are defined by $A^\alpha = (\Phi/c,\,{\bf A})$. For a charge at rest, 
$\tensor*{\dot z}{_\ret^\,^\alpha} = (1,0,0,0)$, and (\ref{phi_ret}) defines $\Phi$ to be the Coulomb potential. 

The electromagnetic field tensor corresponding to (\ref{phi_ret}) is 
\be
          \tensor{F}{_\ret_\,_\alpha_\beta} \,=\, \tensor{A}{_\beta_,_\alpha} - \tensor{A}{_\alpha_,_\beta}\,. 
\ee 
The coordinate derivatives of the potential can be calculated using the relation
\be
           {s_\ret(x)}_{,\beta} \,=\, - \tensor{\zeta}{_\beta}R^{-1} \equiv -\tensor{k}{_\beta}\,, 
\ee 
that follows from differentiating the equation $\zeta^\mu \zeta^\nu \tensor{\eta}{_\mu_\nu} = 0$ 
with respect to $\tensor{x}{^\beta}$. Here we have introduced the null vector $k = \zeta R^{-1}$, 
satisfying the normalization $k \cdot {\dot z}_\ret = -1$. The  Frenet frame components of the field tensor are 
then given by $\tensor*{F}{_\ret^\,^i^j} = \tensor{F}{_\ret_\,_\alpha_\beta} \N{^i^\alpha} \N{^j^\beta}$. One
finds that
\be
\fl  \tensor*{F}{_\ret^\,^i^j} \,=\, \frac{q}{4\pi\epsilon_0 c} 
                       \left( 
                            \left( 
                                   \tensor*{\dot z}{_\ret_\,^\,^i}  \tensor{k}{^j}  
                                  -  \tensor*{\dot z}{_\ret_\,^\,^j} \tensor{k}{^i}
                            \right) \left(1 + aR \right) R^{-2}   
                          + \left( 
                                   \tensor*{\ddot z}{_\ret_\,^\,^i} \tensor{k}{^j}   
                                  - \tensor*{\ddot z}{_\ret_\,^\,^j} \tensor{k}{^i} 
                            \right) R^{-1} 
                       \right),
\label{F_ret}
\ee
where $a$ is the acceleration dependent scalar field
\be
                          a \,=\,  \ddot z_\ret \cdot k\,.
\label{def_a}        
\ee
The frame components of the derivative fields 
$\tensor{\dot z}{_\ret} = \tensor*{\dot z}{_\ret^\,^j}\N{_j}$ and  $\tensor{\ddot z}{_\ret} = \tensor*{\ddot z}{_\ret^\,^j}\N{_j}$ 
can be read from the series expressions (\ref{dot_z_series}) and (\ref{ddot_z_series}) of the following section. Likewise, $\zeta$ is
given by the series (\ref{zetaDelta}) and the scalars $R$ and $a$ then follow from (\ref{def_R}) and (\ref{def_a}). 


\subsection{Tube coordinate expressions for retarded fields}
\label{Retarded fields}


Recall that the retarded proper distance $s_\ret(x)$, is such that $x$ lies on the forward null-cone based at $z(s_\ret(x))$.
The retarded tube coordinate system is constructed so that $s_\ret(x) = s$. In any other 
tube coordinate system one must solve for $s_\ret(x)$ using the condition that $\zeta(x)$, given by (\ref{zeta}), 
is a null vector field. 

For points sufficiently close to ${\cal C}$, the separation between $s$ and $s_\ret(x)$ will be small.
Thus, making use of (\ref{xpos_eps}) and writing $s_\ret(x) = s + \Delta$, one expands (\ref{zeta}) as 
\bea
\fl          \zeta   &=&  r\left(n + \varepsilon \N{_0}\right)
                          - \left( \dot z \,\Delta + \ddot z \,\Delta^2/2! + \dddot z \,\Delta^3/3! \right) + \Or(\Delta^4)
\nonumber \\                          
\fl                  &=& r\left(n + \varepsilon \N{_0}\right)
           - \N{_0}\Delta 
           - \N{_1}\kap{_1}\,\frac{\Delta^2}{2!}
           - \left( \N{_0}\kap{_1}^2 
                  + \N{_1}\tensor{\dot\kappa}{_1}
                  + \N{_2}\kap{_1}\kap{_2} 
             \right)\frac{\Delta^3}{3!}
           + \Or(\Delta^4)\,.
\label{zetaDelta}
\eea
Let $\Delta = \sum_{k=1}^\infty c_k r^k$, for coefficients $c_k=c_k(s,\vartheta,\varphi)$ to be determined. Substituting
the series for $\Delta$ into (\ref{zetaDelta}) gives $\zeta$ as a series in $r$. Then, by forming the series for the null
expression $\zeta\cdot\zeta$ and equating coefficients to zero, one obtains a sequence of equations that can be solved in turn 
for the unknown coefficients $c_k$. The first coefficient is found to be $c_1 = \varepsilon \pm 1$.  
Since the proper distance parameter has been taken to increase to the future, the two solutions for $c_1$ give
$s_\ret(x) = s - r(1-\varepsilon) + \Or(r^2)$ and  $s_\adv(x) = s + r(1+\varepsilon) + \Or(r^2)$. That is, the retarded solution 
corresponds to $c_1 = \varepsilon - 1$. For brevity, let $\mu = \varepsilon - 1$. The first few coefficients in the
series for $\Delta$ are then
\bea
\fl                c_1 &=&  \mu \,,
\quad                
                   c_2 \,=\,  \frac{1}{2}\, \mu^2 \kap{_1}\sin\vartheta\cos\varphi \,, 
\nonumber\\                
\fl                c_3 &=&  \frac{\mu^3}{24} \left( - \left(4 + 3\mu\right) \kap{_1}^2 
                                             +  3 \left(4 + \mu\right) \left(\kap{_1} \cos\varphi \sin\vartheta\right)^2  
                                             +  4  \sin\vartheta \left(\kap{_1}\kap{_2}\sin\varphi  
                                                                 +  \tensor{\dot \kappa}{_1}\cos\varphi \right) \right)   \,.                 
\nonumber
\eea
Observe that in retarded tube coordinates (for which $\mu = \varepsilon - 1 = 0$), one has $c_k=0$ and thus $s_\ret(x) = s$, as expected.

Using the above solution for $s_\ret(x) = s + \Delta$, the fields $\dot z_\ret(x)$ and $\ddot z_\ret(x)$, 
needed in expression (\ref{F_ret}) for the frame components of the retarded electromagnetic field, can be obtained by expansion in $r$. 
One finds that
\bea
\fl        \dot z_\ret(x) &=&  \N{_0} 
                                               + \mu \N{_1} \kap{_1} r 
                                               + \frac{\mu^2}{2}\left( \N{_0}\kap{_1}^2
                                                                     + \N{_1} \!\left(\kap{_1}^2\sin\vartheta\cos\varphi + \tensor{\dot \kappa}{_1}\right)
                                                                     + \N{_2}\kap{_1} \kap{_2} \right)r^2 + \Or(r^3),        
\nonumber \\
\fl              &&
\label{dot_z_series}
\\
\fl        \ddot z_\ret(x) &=& \N{_1} \kap{_1} + \mu \left( \N{_0}\kap{_1}^2
                                                                     + \N{_1}\tensor{\dot \kappa}{_1}
                                                                     + \N{_2}\kap{_1} \kap{_2} \right) r
\nonumber \\ \fl && {}
                         + \frac{\mu^2}{2} \,\Bigl( \N{_0} \kap{_1}\!\left(\kap{_1}^2 \sin\vartheta\cos\varphi + 3\tensor{\dot \kappa}{_1}\right) 
                                                + \N{_1} \!\left(   \kap{_1}^3 
                                                                - \kap{_1} \kap{_2}^2 
                                                                + \kap{_1}\tensor{\dot \kappa}{_1}\sin\vartheta\cos\varphi
                                                                + \tensor{\ddot \kappa}{_1}  \right)
\nonumber \\ \fl && {}
                                                + \N{_2} \!\left( \kap{_1}^2 \kap{_2} \sin\vartheta\cos\varphi
                                                               + 2 \tensor{\dot \kappa}{_1}\kap{_2}
                                                               +   \kap{_1}\tensor{\dot \kappa}{_2}  \right)
                                                + \N{_3} \kap{_1} \kap{_2} \kap{_3}
                                            \Bigr) \,r^2 + \Or(r^3).
\label{ddot_z_series}
\eea
The resulting expressions for $\tensor*{F}{_\ret^\,^i^j}$ can be found in \cite{Norton:08}. 

\subsubsection{Electromagnetic field in retarded tube coordinates}
\label{Retarded fields}
The field $\tensor*{F}{_\ret}$ is particularly simple in retarded tube coordinates because in this case
\be
\fl  \dot z_\ret = \N{_0},\quad 
     \ddot z_\ret = \kap{_1}\N{_1},\quad 
     k = n + \N{_0},\quad
     R = r\,,\quad 
     a = \kap{_1}\sin\vartheta\cos\varphi\,.
\label{ret_vars}
\ee
The frame components of an electromagnetic field tensor are identified with the electric and magnetic fields 
in that frame by
\be
    \left(\tensor{F}{^i^j}\right) \;=\; \left( 
\begin{array}{cccc}
                   0     &  E_1/c  &   E_2/c  &   E_3/c  
\smallskip \\
                  -E_1/c  &    0     &    B_3    &   -B_2  
\smallskip \\
                  -E_2/c  &   -B_3   &     0     &    B_1  
\smallskip \\
                  -E_3/c  &    B_2   &   -B_1    &    0   
\end{array}
\right),
\ee
for example, $\tensor{F}{^0^1} = E_1/c$. The electric and magnetic fields for $\tensor*{F}{_\ret}$ with respect to the 
Frenet frame defined by the retarded tube coordinate system are easily found from (\ref{F_ret}) using (\ref{ret_vars}), 
\bea 
         E \;=\; \frac{q}{4\pi\epsilon_0} \left( \,\frac{n}{r^2} + \left(n \sin\vartheta\cos\varphi - \N{_1} \right) \frac{\kap{_1}}{r} \right),
\\
        B \;=\; \frac{q}{4\pi\epsilon_0 c}\, \biggl( -\N{_2} \cos\vartheta  + \N{_3} \sin\vartheta\sin\varphi \biggr) \,\frac{\kap{_1}}{r}\,.
\label{Bret}
\eea
The magnetic field in the frame associated with time-symmetric tube coordinates
 is non-singular. The singular field (\ref{Bret}) can be traced to mixing of $E$ and $B$ by a boost, of approximate 
magnitude $r \kap{_1}$, that relates these two different frame fields.

\subsection{Expansion of the background field near ${\cal C}$}
\label{background_field}

The values of the background field 
$F = \tensor{F}{^\alpha^\beta}\tensor{\bf e}{_\alpha}\otimes\,\tensor{\bf e}{_\beta} 
   = \tensor{F}{^i^j}\N{_i}\otimes\N{_j}$ on the worldtube cross-sections $\Omega(r,s)$ are to be approximated  
in terms of frame components of the Minkowski derivatives of $F(x)$ at the point $z(s)$ on 
the worldline ${\cal C}$. By Taylor expansion, 
$$
  \tensor{F}{^i^j}(s,r,\vartheta,\varphi) \,=\, \left[ \tensor{F}{^i^j}\right]_{r=0} 
                                                 +  r \left[ \frac{\partial \tensor{F}{^i^j}}{\partial r} \right]_{r=0}
                                                 +  \frac{r^2}{2!} \left[ \frac{\partial^2 \tensor{F}{^i^j}}{\partial r^2} \right]_{r=0}
                                                 +  \ldots
$$
The frame field $\N{_i_\!}(s)$ is independent of $r$, so the radial derivatives can be calculated as 
\bea
       \frac{\partial \tensor{F}{^i^j}}{\partial r} 
             &=& \frac{\partial x^\mu}{\partial r} \frac{\partial }{\partial x^\mu} \left( \tensor{F}{^\alpha^\beta} \right) \N{^i_\alpha}\N{^j_\beta}
\nonumber\\
             &=& \frac{\partial }{\partial r}\left(x^k \N{_k^\mu}\right) \tensor{F}{^\alpha^\beta_{,\mu}} \N{^i_\alpha} \N{^j_\beta}
\nonumber\\
             &=& \frac{\partial x^k}{\partial r} \tensor{F}{^i^j_k} \,, 
\eea
where $x^k$ and $\tensor{F}{^i^j_k}$ are respectively the frame components of the  
Minkowski position vector $x = x^\beta \tensor{\bf e}{_\beta} = x^k \N{_k}$, and the first Minkowski coordinate derivative of $F$. 
Similarly, for example, 
\be
       \frac{\partial^2 \tensor{F}{^i^j}}{\partial r^2} \,=\, \frac{\partial x^k}{\partial r} \frac{\partial x^l}{\partial r}\,\tensor{F}{^i^j_k_l} 
                                                            + \frac{\partial^2 x^k}{\partial r^2}\,\tensor{F}{^i^j_k} \,.
\ee            
From (\ref{xpos_eps}) one sees that $\partial x^k/\partial r = \N{^k}\cdot (n + \varepsilon \N{_0}) = n^k + \varepsilon\,\tensor*{\delta}{^k_0}$, where the 
components of $n(s,\vartheta,\varphi)$ are given by (\ref{n_defn}). The second and higher $r$-derivatives of $x^k$ are zero. The required Taylor 
expansion is therefore
\be
\fl   \tensor{F}{^i^j}(s,r,\vartheta,\varphi) \,=\, \tensor{F}{^i^j^\!}(s) 
              + r \! \left(n^k + \varepsilon\,\tensor*{\delta}{^k_0}\right)  \!\tensor{F}{^i^j_k_\!}(s) 
              + \frac{r^2}{2!} \left(n^k + \varepsilon\,\tensor*{\delta}{^k_0}\right)\! 
                               \left(n^l + \varepsilon\,\tensor*{\delta}{^l_0}\right)\! \tensor{F}{^i^j_k_l_\!}(s) + \ldots 
\ee
where field values on ${\cal C}$ are denoted, for example, as $\tensor{F}{^i^j}(s)  
= \left[\tensor{F}{^i^j}(s,r,\vartheta,\varphi)\right]_{r=0}
= \N{^i_\alpha_\!}(s) \N{^j_\beta_\!}(s) \left[\tensor{F}{^\alpha^\beta}(x)\right]_{x=z(s)}$. 
\section{Conclusion}

The paper describes a tube method for calculating the self-force on an extended radiating system.
The new method is presented for the special case of electromagnetic self-force in Minkowski spacetime, but it seems likely
that similar techniques could be used to derive a tube method applicable to a radiating system of finite size in 
a curved background spacetime. 
In particular, a finite radius tube method may offer a mathematically sound alternative to having to 
invoke a mass renormalization in the tube-based gravitational self-force calculations of \cite{Mino:97}.
In contrast to the classical mass renormalisation procedure, all steps in our tube 
method are physically well motivated and could be justified to any required degree of mathematical rigour. 

For finite size electromagnetic systems in Minkowski spacetime, we find that if the field 
outside the radiating system is known exactly (for example, as a prescribed multipole expansion) then 
our tube method gives an exact expression for the self-force. This self-force 
coincides with that derived by the classical mass renormalisation procedure, so in this sense, it has been proved 
that the renormalisation procedure gives correct results in Minkowski spacetime. 
However, as discussed in Section~\ref{cmr}, the reason why classical mass renormalisation works is that the point 
particle limit (tube radius $r \rightarrow 0$)
becomes superfluous if renormalisation is implemented in full using the Bhabha and Harish-Chandra 
inflow theorem. In fact, a fully implemented renormalisation scheme would be exactly equivalent to
our finite radius tube method. Thus, taking $r \rightarrow 0$ only serves to introduce mathematical inconsistencies
and obscure the underlying physics. The classical mass renormalisation procedure is therefore conceptually flawed, in so 
far as it is associated with a point particle limit.

The finite radius tube method follows from Dirac's balance relation together with our expression (\ref{PcapSeries}) 
for the momentum 
flux through the cap of the worldtube. Assuming the weak energy condition for the total energy momentum tensor $T,$ 
expression (\ref{PcapSeries})
can also be used to determine the smallest possible size of an extended radiating system that has an asymptotic field 
with a given multipole structure.
In Section~\ref{sanity_ck} this was done for a particle with an asymptotic field equal to that of an electric monopole. 
We found that if such a particle is governed by the Lorentz--Dirac equation, and if its mass and charge are set equal to that of 
the electron, then this particle is at least a thousand times larger than the length scale of any possible electron structure, 
else the weak energy condition is violated. On the other hand, the experimental limits on electron structure size 
are consistent with the weak energy condition as applied to helical worldline models of a classical spinning electron.

\ack

This work was supported by the Australian Research Council through CUDOS, an ARC 
Centre of Excellence, and through ARC grant DP0559788 at Monash University.  



\end{document}